\documentclass[preprint,12pt]{elsarticle}
\usepackage{graphicx}
\usepackage{amssymb}
\journal{Physica B}
\begin{document}
\title
{Double exchange model on triangular lattice: non-coplanar spin configuration and phase transition near quarter filling}
\author{G. P. Zhang}
\address{Department of Physics, Renmin University of China, Beijing 100872, China}
\ead{zhanggp96@ruc.edu.cn}
\author{Jian Zhang}
\address{3M Company, 3M Corporate Headquarters, 3M Center, St. Paul, MN 55144-1000, United States}
\author{Qi-Li Zhang}
\address{Data Center for High Energy Density Physics,Institute of Applied Physics and Computational Mathematics, Beijing 100094, China}
\author{Jiang-Tao Zhou}
\address{College of Optoelectronic Engineering, Shenzhen University, Shenzhen 518060, China}
\author{M. H. Shangguan}
\address{Department of Physics, Renmin University of China, Beijing 100872, China}

\date{\today}

\begin{abstract}
Unconventional anomalous Hall effect in frustrated pyrochlore oxides is originated from spin chirality of non-coplanar localized spins, which can also be induced by the competition between ferromagnetic (FM) double exchange interaction $J_{H}$ and antiferromagnetic superexchange interaction $J_{AF}$. Here truncated polynomial expansion method and Monte Carlo simulation are adopted to investigate the above model on two-dimensional triangular lattice. We discuss the influence of the range of FM-type spin-spin correlation and strong electron-spin correlation on the truncation error of spin-spin correlation near quarter filling. Two peaks of the probability distribution of spin-spin correlation in non-coplanar spin configuration clearly show that non-coplanar spin configuration is an intermediate phase between FM and 120-degree spin phase. Near quarter filling, there is a phase transition from FM into non-coplanar and further into 120-degree spin phase when $J_{AF}$ continually increases. Finally the effect of temperature on magnetic structure is discussed.

\end{abstract}



\maketitle

\section{Introduction}
Recently unconventional anomalous Hall effect (UAHE) was observed in frustrated pyrochlore oxides Nd2Mo2O7 \cite{2-Science-2001}, Pr2Ir2O7 \cite{5-PRL-2007,8-Nature-2010} and PdCrO2 \cite{11-PRL-2010}, different from that induced by spin-orbit coupling in ordinary magnetic conductors. UAHE is originated from Berry phase in electron hopping amplitude as a result of noncoplanar alignment of localized spin. In frustrated pyrochlore lattices, both the geometric and magnetic frustrations result in noncoplanar configuration of localized spin, i.e., spin chirality. On the side of theoretical research, magnetic phase diagram \cite{1-PRB-2003}, electronic phase separation \cite{9-PRB-2010} and phase competition \cite{16-PRL-2010} had been intensively investigated for double exchange model on frustrated pyrochlore lattice. The magnetic phase diagram is rather insensitive to the ferromagnetic exchange interaction but highly depends on doping \cite{1-PRB-2003}, which also validates for double exchange model in square lattice and triangular lattice.

Meanwhile, more and more attentions were paid to spin chirality in FM Kondo lattice model on two-dimensional triangular lattice \cite{6-PRL-2008,10-JPSJ-2010,kumar-01,akagi-01,zhang}, and noncoplanar spin configurations are stable in a quite large parameters space \cite{akagi-01,12-PRL-2010}. In noncoplanar spin configuration, the density of states and the resistivity at quarter filling are independent of the temperature, while the system undergoes a metal-insulator transition when $J_{AF}$ increases \cite{kumar-01}. Furthermore, the bad metallic behavior is consistent with experimental observation in frustrated itinerant magnets R2Mo2O7 \cite{kumar-01}.
120-degree spin configuration is easily destabilized by electron doping from band bottom and a noncoplanar three-sublattice ordering occurs accompanied by an intervening phase separation, while a noncoplanar ordering does not occur by hole doping from band top \cite{akagi-01}.

On the other hand, due to huge potential application of colossal magnetoresistance effect \cite{CMR1,CMR2}, doped manganese oxide is one intensive research subject. It was found that phase separation \cite{phaseseparation,mcs1} is a key factor for CMR. Electronic phase separation had been verified by various experimental observations \cite{PS1,PS2,PS3}. By dynamic mean-field theory \cite{Tong}, it was found that phase separation between FM and paramagnetic phase and that between FM and AF phase is robust against Coulomb interaction. For double exchange model in two-dimensional triangular lattice, there exists a phase separation from the electronic density in lightly doped region \cite{akagi-01}.

In our previous paper \cite{zhang}, we focused on the application of TPEM to double exchange model on 2D triangular lattice and discussed the spin-spin correlation and spin structure factors in low-, mediate- and high electron density. In this Letter, we still adopt TPEM and Monte Carlo simulation to investigate double exchange model on 2D triangular lattice. We discuss the influence of the range of FM-type spin-spin correlation and strong electron-spin correlation on the truncation error of spin-spin correlation by TPEM near quarter filling. We clearly illustrate non-coplanar spin configuration in one triangle and plot its probability distribution of spin-spin correlation, which has two peaks located at 1 and -0.5 and shows that it is an intermediate phase between FM and 120-degree spin phase. Near quarter filling, the phase transition from FM to non-coplanar phase to 120-degree spin phase occurs when AF SE interaction continually increases. Finally we discuss the effect of temperature on magnetic structure.

\section{Model and method}
The hamiltonian consists of two parts, i.e., $H=H_{DE}+H_{AF}$. $H_{DE}$ describes
electron hopping between nearest-neighboring (NN) lattices and the Hund's-rule interaction between itinerant electron and localized spins,
\begin{equation}
H_{DE}=-t\sum_{<ij>,\alpha}(C_{i,\alpha}^{\dag}C_{j,\alpha}+h_{.}c_{.})
-J_{H}\sum_{i,\alpha,\beta}C_{i,\alpha}^{\dag}\sigma_{\alpha\beta}C_{i,\beta}\cdot S_{i}.
\end{equation}
Here NN hopping integral $t$ is chosen as the energy unit and $J_H$ is the Hund
interaction strength. $C_{i,\alpha}^{\dag}$ ($C_{i,\alpha}$) creates (annihilates) one electron at
the lattice $i$ with the spin orientation $\alpha$, $<ij>$ stands for two nearest-neighboring lattices, and $\sigma_{\alpha\beta}$
is the Pauli matrix.
$H_{AF}=J_{AF}\sum_{<ij>}S_{i} \cdot S_{j}$
describes AF SE interaction between two nearest-neighboring localized spins. All localized spins $S_i$ are assumed as $1$ and treated as a classic field. Electron degree of freedom can be integrated for any given localized spin configuration. Classic Monte Carlo simulation combined with TPEM \cite{TPEM1,TPEM2} is adopted to
investigate the energy, electron density, spin-spin correlation and spin structure factor.
There are three key parameters for
TPEM, i.e., $M$, $\epsilon_{p}$ and $\epsilon_{tr}$, which control the accuracy and computational
speed \cite{zhang,TPEM1,TPEM2}. The maximal Monte Carlo step is 40000, and the physical quantity
is evaluated every 20 steps after first 6000 warmup steps.

\begin{figure}[htb]
\centering
\includegraphics[width=8cm]{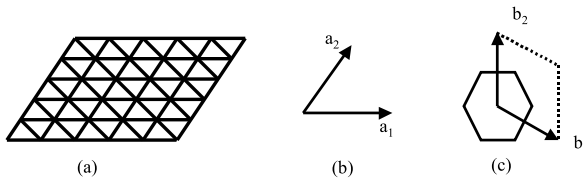}
\caption{ (a) $6 \times 6$ triangular lattice, (b) the lattice vectors and (c) the reciprocal vectors and the
first Brillouin zone.}\label{fig0}
\end{figure}

Fig. \ref{fig0}(a) show $6 \times 6$ triangular lattice with periodic boundary condition. Each lattice has six nearest-neighbors. As shown in Fig. \ref{fig0}(b), two lattice vectors are $\vec{a}_1=(1,0)a$ and $\vec{a}_2=(\frac{1}{2},\frac{\sqrt{3}}{2})a$ with $a$ being the lattice constant. The reciprocal lattice vectors, i.e., $\vec{b}_1=\frac{4\pi}{\sqrt{3}a}(\frac{\sqrt{3}}{2},-\frac{1}{2})$
and $\vec{b}_2=\frac{4\pi}{\sqrt{3}a}(0,1)$, and the first Brillouin zone are shown in Fig. \ref{fig0}(c).
On $L \times L$ triangular lattice, the momentum $\vec{q}$ is $\frac{m}{L}\vec{b}_1+\frac{n}{L}\vec{b}_2$ and shortened as $(q_1,q_2)$ with $q_1$ and $q_2$ being $\frac{m}{L}$ and $\frac{n}{L}$ respectively. $m$ and $n$ are integers within $[0, L]$. Spin structure factor $S(q)$ is defined by
\begin{equation}
S(\vec{q})=\frac{1}{N}\sum_{i,j}<S_i \cdot S_j> e^{i\vec{q} \cdot
\vec{r}_{ij}}.
\end{equation}
$<S_i \cdot S_j>$ is the average spin-spin correlation between the localized spins $S_{i}$ and $S_{j}$, and $\vec{r}_{ij}$ is the displacement from the lattice $i$ to the lattice $j$. Due to periodic boundary condition adopted in our study, only $<S_{1}\cdot S_{j}>$ is shown for simplicity. $j=i_{x}+(i_{y}-1)*L$ stands for the lattice with indices $(i_{x},i_{y})$, in which $1\le i_{x} \le L$ and $1\le i_{y} \le L$, and $N=L \times L$.

\begin{figure}[htb]
\centering
\includegraphics[width=6cm]{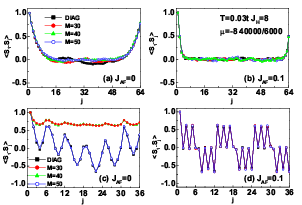}
\includegraphics[width=6cm]{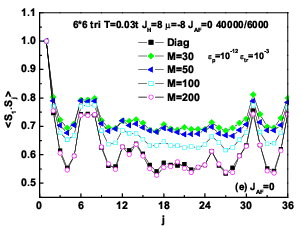}
\includegraphics[width=6cm]{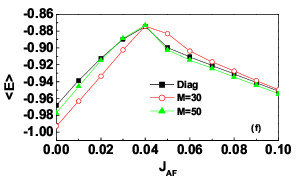}
\caption{The comparison of spin-spin correlation $<S_1 \cdot S_j>$ between exact diagonalization (ED)
and TPEM in (a,b) one-dimensional chain with $L=64$, (c,d) two-dimensional $6 \times 6$ square lattices
and (e) $6 \times 6$ triangular lattice. $J_{AF}=0$ in (a,c,e) and $J_{AF}=0.1$ in (b,d). The case for
$J_{AF}=0.1$ in $6 \times 6$ triangular lattice had been shown in Fig. 4(e) \cite{zhang}.
(f) The relation of the energy $E$ to the strength of antiferromagnetic superexchange interaction $J_{AF}$
calculated by ED and TPEM with $M=30$ and $M=50$. Other parameters are $T=0.03t$, $J_{H}=8$, $\mu=-8$, and $\epsilon_{tr}=10^{-3}$. (a-d,f) $\epsilon_{p}=10^{-5}$ and (e) $\epsilon_{p}=10^{-12}$.}\label{figaccuracy}
\end{figure}

It is well known that the truncation error of the physical quantity decreases when the truncation moment $M$ of TPEM increases, and eventually all physical quantity can be exactly reproduced as $M$ approaches infinity. In 1D and 2D regular lattice with periodic boundary condition, at quarter filling there is no stable FM phase and FM spin-spin correlation is short-ranged in the absence of AF SE interaction. In this case, $M \simeq 30$ and $M \simeq 50$ is enough to reproduce the spin-spin correlation $<S_{1}\cdot S_{j}>$, as shown in Fig. \ref{figaccuracy}(a,c). In the presence of strong AF SE interaction, e.g., $J_{AF}=0.1$, the short-range spin-spin correlation is FM and may becomes AF. In this case, $M=30$ is sufficient to reproduce $<S_{1}\cdot S_{j}>$, as shown in Fig. \ref{figaccuracy}(b,d) and Fig. 4(e) in Ref. \cite{zhang}. It was found that a spin-flux phase occurs on 2D square lattice {\cite{TPEMappl1,spin-flux}}, where four localized spins within each square lie (anti)clockwise at the same plane. However, a new type of spin-flux phase, i.e., noncoplanar spin configuration, occurs on 2D triangular lattice, as a result of the competition of electron-mediated FM alignment and AF SE interaction \cite{kumar-01,akagi-01,zhang}. However, $M \ge 200$ is required to reproduce FM-type spin-spin correlation $<S_{1}\cdot S_{j}>$ at quarter filling in 2D triangular lattice \cite{zhang}, as shown in Fig. \ref{figaccuracy}(e). These results imply that the range of FM-type spin-spin correlation and thus FM magnetic domain is an important factor for the accuracy of spin-spin correlation by TPEM. In constrast, at a low truncation moment $M$ and a large parameters space, most physical quantities such as the energy $E$ and the electron density $n$ are accurate compared with the exact value. As shown in Fig. \ref{figaccuracy}(f), the energy can be well reproduced by TPEM at $M \simeq 50$, irrelevant of the strength of AF SE interaction $J_{AF}$.

On the other hand, we explore the effect of electron mediation on the accuracy of spin-spin
correlation by TPEM. In three-dimensional (3D) cube lattice, the magnetization at
quarter filling is accurate at $M=30$ \cite{TPEM2}, compared with exact diagonalization method.
In 2D triangular lattice, $M=30$ is adequate to reproduce spin-spin correlation at low electron filling (e.g., $n$=0.1944), even in the presence of long-range FM-type spin-spin correlation \cite{zhang}. The electron-spin correlation in the above cases is weaker than that in triangular lattice near quarter filling. Therefore strong electron mediation effect, i.e., strong electron-spin correlation, is the other important factor for the accuracy of spin-spin correlation by TPEM.

\section{Result and discussion}
\subsection{Non-coplanar Spin configuration in one triangle}
\begin{figure}[htb]
\centering
\includegraphics[width=8cm]{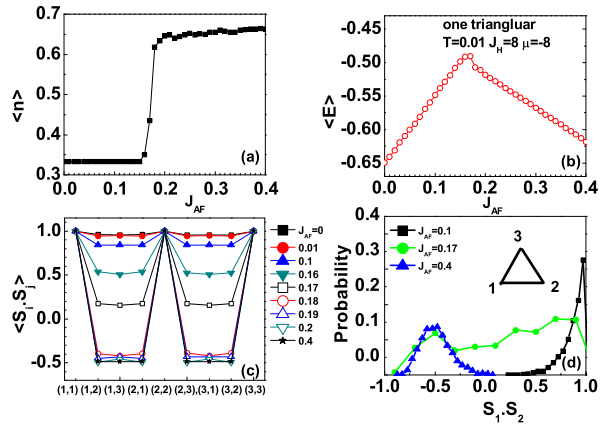}
\caption{In one triangular lattice, (a) the electron density $n$, (b) the energy $E$, (c) the spin-spin correlation $<S_i \cdot S_j>$ and (d) the probability distribution of $S_1 \cdot S_2$ as $J_{AF}$ increases.
Other parameters are $T=0.01t$, $J_{H}=8$ and $\mu=-8$.}\label{fig3-onetriangular}
\end{figure}

Due to simplicity, one triangular lattice is chosen to show how the spin configuration changes with
AF superexchange interaction. The possible electron density $n$ of one triangular lattice is 0, 1/3, 2/3 and 1.
It is found that the localized spins are 120$^{\circ}$ configuration as $n$ is $\frac{2}{3}$ and 1, which is originated from the geometrical frustration and independent of the value of $J_{AF}$ .
Fig. \ref{fig3-onetriangular}(a-c) plots the changes in the electron density $n$, the energy $E$ and spin-spin correlation as $J_{AF}$ increases from 0 to 0.4. It is found that $n$ is 1/3 and 2/3 correspondingly as $J_{AF}$ is weaker than 0.17 and stronger than 0.18 respectively, and $n$ sharply changes from 1/3 to 2/3 as $J_{AF}$ is 0.17$\sim$0.18. As $J_{AF}$ increases, $E$ first increases till reaches a maximum -0.49071 and then monotonically decreases. The spins are parallel at $n=1/3$ and the spin-spin correlation is 1 in the presence of a weak AF SE interaction. As $J_{AF}$ increases, the spin configuration evolves from parallel into non-coplanar configuration and further into 120$^{\circ}$ configuration, and the spin-spin correlation between two nearest-neighboring localized spins changes from 1 to -0.5. Interestingly, the maximal energy $E_{max}$ occurs at the transition point of the electron density and the spin-spin correlation. Fig. \ref{fig3-onetriangular}(d) shows the probability distribution of $S_{1}\cdot S_{2}$ as $J_{AF}=0.1$, 0.17 and 0.4. The corresponding mean value, i.e., $\langle S_{1}\cdot S_{2}\rangle$, is 0.8381, 0.1753 and -0.48842 respectively. It clearly shows that the peak of probability distribution is located at 1 (for FM phase) and -0.5 (for 120$^{\circ}$-configuration). Surprisingly, the probability distribution has two peaks that are located at 1 and -0.5 respectively for non-coplanar spin configuration. In one triangular lattice, since $n=2/3$ induces 120$^{\circ}$ spin configuration, non-coplanar spin configuration only exists in a smaller space of parameters compared with a large system.

\subsection{Effect of AF superexchange interaction on spin configuration}
\begin{figure}[htb]
\centering
\includegraphics[width=6cm]{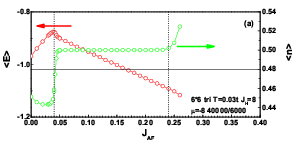}
\includegraphics[width=6cm]{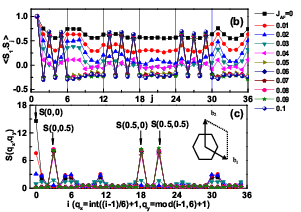}
\caption{(a) The energy $E$ and the electron density $n$, (b) the spin-spin correlation $<S_1 \cdot S_j>$
and (c) the spin structure factor $S(q_{x},q_{y})$ when $J_{AF}$ changes from $0$ to $0.1$. $T=0.03t$,
$J_{H}=8$ and $\mu=-8$.}\label{fig3}
\end{figure}

In a large triangular lattice, we further investigate the effect of AF superexchange interaction on the energy $E$ and the spin configuration by continuously tuning the strength of AF superexchange interaction $J_{AF}$. As $J_{AF}$ increases from $0$ to $0.038t$, $E$ increases, as shown in Fig. \ref{fig3}(a).
The increase of $E$ with $J_{AF}$ is originated from the fact that the spin configuration is in FM phase at a weak $J_{AF}$, therefore the spin-spin correlation between every two nearest-neighboring localized spins $\langle S_{i}\cdot S_{j}\rangle$ is positive. The FM phase is strengthened by the six nearest neighboring lattices surrounding each lattice in 2D triangular lattice \cite{zhang}, compared with unstable FM phase in 2D square lattice. When $J_{AF}$ further increases, FM-like spin-spin correlation decreases and the angle between nearest neighboring localized spins approaches to 90$^{\circ}$,
as shown in Fig. \ref{fig3}(b). As a result, the spin structure factor at the vector $(0,0)$, i.e.
$S(0,0)$, decreases as $J_{AF}$ increases, as shown in Fig. \ref{fig3}(c).
Since $\sum_{<ij>}\langle S_{i}\cdot S_{j} \rangle$ remains positive, the energy $E$ increases
with $J_{AF}$ by a competition between increasing $J_{AF}$ and decreasing $\langle S_{i}\cdot S_{j} \rangle$.

As $J_{AF}$ is larger than a critical value $J_{AF}^{c1}=0.038t$, the energy $E$ decreases from a maximum as shown in Fig. \ref{fig3}(a). Therefore the sign of the slope of the energy in term of $J_{AF}$ changes from positive to negative, which indicates a phase transition \cite{zhangQL}. As $J_{AF}$ is strong, the energy corresponding to FM phase is too large and a new phase with a lower energy emerges, i.e., noncoplanar spin configuration.
In this phase, the spin-spin correlation between any two nearest neighboring localized spins is negative,
as shown in Fig. \ref{fig3}(b), while $(S_i \times S_j)\cdot S_k$ with the sites $i$, $j$ and $k$
belonging to one triangle is nonzero \cite{zhang}. Noncoplanar spin configuration is different from both FM phase and 120$^{\circ}$ configuration phase. The spin structure factor $S(\vec{q})$ has three peaks, which
are located at the vectors $(0.5, 0)$, $(0, 0.5)$ and $(0.5, 0.5)$ respectively \cite{kumar-01}, as shown in
Fig. \ref{fig3}(c). Rescaled spin structure factor $S(\vec{q})/N \sim 0.29$ is close to its maximum value 1/3, and finite size effect is small \cite{zhang}.

When $J_{AF}>J_{AF}^{c2}=0.225t$, the system eventually evolves into 120$^{\circ}$ spin configuration phase.
In this phase, the spin-spin correlation between any two nearest-neighboring localized spins closes to $-0.5$,
as shown in corrected Fig.4(a,d) \cite{zhang}, while $(S_i \times S_j)\cdot S_k$ with the sites $i$, $j$ and $k$
belonging to one triangle is zero \cite{zhang}. The spin structure factor $S(\vec{q})$ has two peaks, which
is located at vectors $({1}/{3}, {2}/{3})$ and $({2}/{3}, {1}/{3})$ respectively \cite{kumar-01}, as shown in corrected Fig. 5(a,d) \cite{zhang}.
Rescaled spin structure factor $S(\vec{q})/N$ is close to its maximum value 1/2 \cite{kumar-01,zhang}.
The phase transition from noncoplanar spin configuration to 120-degree spin phase, which is consistent with the phase diagram shown in Ref. \cite{kumar-01} in the limit of infinite Hund interaction.

Finally we illustrate how the electron density $n$ changes with $J_{AF}$ in Fig. \ref{fig3}(a). Different from 2D square lattices, $n$ does not always equal to 0.5 for 2D triangular lattice as $\mu=-J_{H}$. Of course, one can fix $n$ at 0.5 by adjusting the chemical potential $\mu$. However, it is complicated to compare the energy when both $\mu$ and $J_{AF}$ varies. Since the deviation of $n$ from 0.5 does not exceed $0.06$ for FM phase and 120$^{\circ}$ configuration, we still set $\mu$ as -8 in our investigation. Interestingly, $n$ equals to 0.5 when the system is in non-coplanar spin configuration, as shown in Fig. \ref{fig3}(a).

\subsection{Effect of temperature on spin configuration}

\begin{figure}[htb]
\centering
\includegraphics[width=6cm]{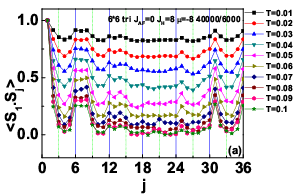}
\includegraphics[width=6cm]{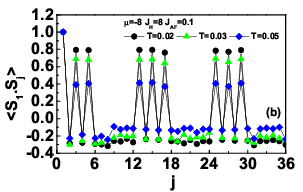}
\includegraphics[width=6cm]{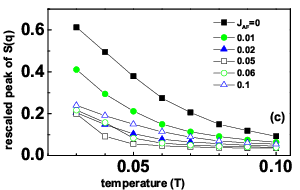}
\caption{The temperature dependence of (a) spin-spin correlation $<S_1 \cdot S_j>$ at $J_{AF}=0$, (b) $<S_1 \cdot S_j>$ at $J_{AF}=0.1$ and (c) the rescaled peak of $S(q_{x},q_{y})$.}\label{fig4}
\end{figure}

We show the effect of temperature on the spin-spin correlation at $J_{AF}=0$ and 0.1 respectively in Fig. \ref{fig4}(a) and (b). As shown in Fig. \ref{fig4}(a), FM-type spin-spin correlation at $J_{AF}=0$ decreases as the temperature $T$ increases, and the system tends to be paramagnetic as the temperature is higher than 0.1$t$.
On the other hand, the absolute value of the spin-spin correlation of non-coplanar spin configuration also decreases
as $T$ increases, as shown in Fig. \ref{fig4}(b).
When $J_{AF}$ increases from 0 to 0.1, the peak of spin structure factor moves shifts from the vector $(0,0)$ to the vectors $(0,0.5)$, $(0.5,0)$ and $(0.5,0.5)$, by accompanying the reduction of $S(0,0)$ and the increase of $S(0,0.5)$. The rescaled peak of the spin structure factor $S(q_{x},q_{y})$ decreases when the temperature $T$ increases,
as shown in Fig. \ref{fig4}(c). All rescaled peaks collapse at $T=0.1$, which implies the system is in
paramagnetic phase. The temperature dependence of spin-spin correlation and no phase transition at high temperature is consistent with the phase diagram shown in Ref. \cite{kumar-01}.

\section{Conclusion}

We investigate the effects of AF SE interaction and temperature on the spin configuration near quarter filling for double exchange model on triangular lattice. Non-coplanar spin configuration is induced by competition between FM parallel alignment mediated by itinerant electrons and AF SE interaction between localized spins. A phase transition from FM into non-coplanar phase and further into 120-degree spin phase occurs when AF SE interaction becomes stronger. At high temperature, there is no phase transition and all magnetic phase is paramagnetic.

\section*{Acknowledgements} G. P. Zhang thanks Prof. X-Q Wang
for proposing this interesting project on double exchange model in triangular lattice
and many insightful discussions. G. P. Zhang also thanks Dr. N. H. Tong for his critical reading of this manuscript and insightful suggestions. This work was supported by the National Basic Research Program
of China (2012CB921704) and the National Natural Science Foundation of China under Grant No 11174363.

\end{document}